\documentclass[twocolumn,tighten]{aastex63}

\usepackage{graphicx}
\usepackage{amsmath}
\usepackage{amssymb}
\usepackage{color}
\usepackage{mathtools}
\usepackage{ulem}
\usepackage[all]{hypcap}

\newcommand\msun{\, {M}_\odot}

\newcommand\kms{\, {\rm km}\,{\rm s}^{-1}}

\newcommand\gpcyr{\, {\rm Gpc}^{-3}\,{\rm yr}^{-1}}

\newcommand\mseed{M_{\rm seed}}

\newcommand\vesc{{v_{\rm esc}}}

\def\apjl{ApJL}%
\def\apjs{ApJS}%
\def\aap{A\&A}%
\begin{document}
\shorttitle{IMBH mergers in NSCs}
\shortauthors{Fragione et al.}

\title{Merger rates of intermediate-mass black hole binaries in nuclear star clusters}

\correspondingauthor{Giacomo Fragione}
\email{giacomo.fragione@northwestern.edu}

\author[0000-0002-7330-027X]{Giacomo Fragione}
\affil{Center for Interdisciplinary Exploration \& Research in Astrophysics (CIERA) and Department of Physics \& Astronomy, Northwestern University, Evanston, IL 60208, USA}
\affil{Department of Physics \& Astronomy, Northwestern University, Evanston, IL 60202, USA}

\author[0000-0003-4330-287X]{Abraham Loeb}
\affil{Astronomy Department, Harvard University, 60 Garden St., Cambridge, MA 02138, USA}

\author[0000-0002-4865-7517]{Bence Kocsis}
\affil{Rudolf Peierls Centre for Theoretical Physics, Clarendon Laboratory, Parks Road, Oxford OX1 3PU, UK}

\author[0000-0002-7132-418X]{Frederic A. Rasio}
\affil{Center for Interdisciplinary Exploration \& Research in Astrophysics (CIERA) and Department of Physics \& Astronomy, Northwestern University, Evanston, IL 60208, USA}
\affil{Department of Physics \& Astronomy, Northwestern University, Evanston, IL 60202, USA}

\begin{abstract}
Repeated mergers of stellar-mass black holes (BHs) in dense star clusters can produce intermediate-mass black holes (IMBHs). In particular, nuclear star clusters at the centers of galaxies have deep enough potential wells to retain most of the BH merger products, in spite of the significant recoil kicks due to anisotropic emission of gravitational radiation. These events can be detected in gravitational waves (GWs), which represent an unprecedented opportunity to reveal IMBHs. In this paper, we analyze the statistical results of a wide range of numerical simulations, which encompass different cluster metallicities, initial BH seed masses, and initial BH spins, and we compute the merger rate of IMBH binaries. We find that merger rates are in the range $0.01$-$10$\,Gpc$^{-3}$\,yr$^{-1}$ depending on IMBH masses. We also compute the number of multi-band detections in ground-based and space-based observatories. Our model predicts that a few merger events per year should be detectable with LISA, DECIGO, ET, and LIGO for IMBHs with masses $\lesssim 1000\msun$, and a few tens of merger events per year with DECIGO, ET, and LIGO only.
\end{abstract}

\section{Introduction}
\label{sect:intro}

Astrophysical black holes (BHs) are classified according to their mass as stellar-mass BHs ($\lesssim 10^2\msun$), intermediate-mass black holes (IMBHs; $\sim 10^2\msun - 10^5\msun$), or supermassive BHs ($\gtrsim 10^5\msun$). IMBHs could play a fundamental role in a wide variety of contexts: they could be the seeds that grow into supermassive BHs at the centers of galaxies, providing feedback on galaxy evolution and a possible source of reionization \citep[e.g.,][]{madau2001,Silk2017,tagawa2019,Natarajan2021}; they can produce observable tidal disruption events \citep[e.g.,][]{RosswogRamirez-Ruiz2009,chen2011,MacLeodGuillochon2016,fragle2018}; they can for binaries that become gravitational wave (GW) sources with unique properties \citep[e.g.,][]{Miller2002,MandelBrown2008,gair2011,fragl2018b,ArcaSeddaAmaroSeoane2021}; and they could be accreting from a stellar binary companion, producing ultra-luminous X-ray sources \citep[e.g.,][]{kaaret2017ARA&A..55..303K}. Therefore, characterizing the population of IMBHs represents a crucial step towards our understanding of the Universe \citep*[see][for a review]{GreeneStrader2020}. 

There are four main observational methods to detect IMBHs. The first two consist of tracking stellar and gas dynamics and modeling gas accretion around a massive BH, respectively. Both methods have provided several massive ($\sim 10^4 - 10^5\msun$) nearby candidates \citep[e.g.,][]{BaldassareReines2015,chili2018,PechettiSeth2022,MicicIrwin2022}, but none of them has confirmed the existence of an IMBH beyond reasonable doubt. The third method is based on looking for tidal disruption events consistent with an IMBH, similarly to what is typically done for supermassive BHs in galactic nuclei \citep[e.g.,][]{lin2018,peng2019,shen2019}. Detecting a signal in the outskirts of a galaxy and from the disruption of a white dwarf would be a promising evidence for the existence of an IMBH \citep[e.g.,][]{rossw2008,RosswogRamirez-Ruiz2009,MacLeodGuillochon2016}, and new facilities, such as JWST and LSST/VRO, may be able to detect tens of these events out to redshift $z\sim 1$. The fourth method consists of looking for GW events, where one of the components in the merging binary is in the IMBH mass range. While LIGO/Virgo/KAGRA can detect the mergers of IMBHs with masses $\sim 100\,M_\odot$ out to $z\sim 1$ \citep[e.g.,][]{AbbottAbbott2019}, as already done in the notable case of GW190521 \citep{AbbottAbbott1905212020}, the upcoming LISA, ET, TianQin, and DECIGO have the potential to detect IMBHs throughout the Universe \citep[e.g.,][]{LuoChen2016,Amaro-SeoaneAudley2017,JaniShoemaker2020}. Current LIGO/Virgo/KAGRA upper limits for mergers of IMBHs with masses up to about $500\msun$ are $\sim 0.1-10\gpcyr$ \citep{AbbottAbbott2017imbh,AbbottAbbott2019,AbbottAbbott2022imbh}.

Three main pathways to form IMBHs have been discussed in the literature. The first two channels involve the direct collapse of a gas cloud of pristine gas \citep{loeb1994,bromm2003,begelm2006} or of massive Pop III stars \citep{madau2001,bromm2004,fryer2001} at high redshift, which might form an IMBH of $\sim 10^4 - 10^5\msun$ and $\sim 100\msun$, respectively. IMBHs with masses in between these two extremes can be produced at any redshift through repeated mergers either of massive main-sequence stars, later collapsing to form a BH, \citep{por02,gurk2004,frei2006,panloeb2012,gie15,TagawaHaiman2020,DasSchleicher2021,DiCarloMapelli2021}, or of stellar-mass BHs \citep{mil02b,OLeary+2006,antoras2016,antonini2019,FragioneKocsis2022,GonzalezKremer2021,MapelliDall'Amico2021,WeatherfordFragione2021}. Other possibilities include the fragmentation of AGN disks \citep{McKernan+2012,McKernan+2014}, super-Eddington accretion onto stellar BHs embedded in AGN disks \citep[e.g.,][]{koc11}, and repeated mergers of BHs with stars in galactic nuclei \citep[e.g.,][]{StoneKupper2017,RoseNaoz2021}. 

GWs provide the most secure way to weigh IMBHs since their mass is strictly encoded in the chirp mass and the characteristic strain of the signal. Only a few studies systematically explore the growth and formation of IMBHs in the context of repeated mergers \citep[e.g.,][]{antonini2019,FragioneKocsis2022,MapelliDall'Amico2021}, and even fewer predict their merger rates to compare to present and future constraints from GW detectors \citep[e.g.,][]{fragk18,fragleiginkoc18,RasskazovFragione2020}. However, IMBH merger rates are typically predicted in the range $\sim 0.01-10\gpcyr$, which would imply several detectable mergers per year with current and upcoming detectors. In this paper, we use the results from a large set of semi-analytic calculations, performed with the code developed in \citet{FragioneKocsis2022}, to study the formation of IMBHs via repeated mergers, and we compute for the first time the cosmic rate of IMBH mergers in nuclear star clusters (NSCs). We present merger rates as a function of the initial seed mass and the characteristic BH spin at birth, and we make predictions about the multi-band detectability of mergers of IMBH binaries.

This paper is organized as follows. In Section~\ref{sect:method}, we discuss our numerical framework to study the formation and mergers of IMBHs. In Section~\ref{sect:results}, we present and discuss our results for merger rates and multi-band detectability. Finally, in Section~\ref{sect:concl}, we discuss the implications of our results and draw our conclusions.

\section{Method}
\label{sect:method}

We consider NSCs since they represent the ideal environment to form IMBHs via repeated mergers owing to their large escape speeds \citep[e.g.,][]{antonini2019,FragioneKocsis2022,MapelliDall'Amico2021}. In what follows, we summarize the semi-analytical scheme developed in \citet{FragioneKocsis2022} that we adopt here.

We assume that the NSC density is described by a three-parameter potential-density pair \citep[for details see][]{StoneOstriker2015}. To generate a population of NSCs, we start from sampling galaxy masses ($M_{\rm *,gal}$) from a Schechter function
\begin{equation}
    \Phi(M_{\rm *,gal}) \propto \left(\frac{M_{\rm *,gal}}{M_{\rm c}} \right)^{-\alpha_{\rm c}} \exp\left(-\frac{M_{\rm *,gal}}{M_{\rm c}}\right)\,,
\end{equation}
where $M_{\rm c} = 10^{11.14}\msun$ and $\alpha_{\rm c}=1.43$ \citep{FurlongBower2015}. To scale the galaxy mass to the NSC mass ($M_{\rm NSC}$), we use scaling relations for late-type galaxies from \citet{georg2016}, 
\begin{equation}
    \log(M_{\rm NSC}/c_1)=\zeta\times\log(M_{\rm *, gal}/c_2)+\psi\,,
    \label{eqn:mclmgal}
\end{equation}
and, to sample their half-mass radius ($r_{\rm h}$), we use
\begin{equation}
    \log(r_{\rm h}/c_3)=\kappa\times\log(M_{\rm NSC}/c_4)+\omega\,,
    \label{eqn:rhmcl}
\end{equation}
where $c_1= 2.78\times 10 ^6\msun$, $c_2= 3.94\times 10^9\msun$, $\zeta = 1.001^{+0.054}_{-0.067}$, $\psi=0.016^{+0.023}_{-0.061}$ and $c_3=3.31$~pc, $c_4=3.60\times 10^6\msun$, $\kappa=0.321^{+0.047}_{-0.038}$, $\omega=-0.011^{+0.014}_{-0.031}$. Note that we consider the scatter in the fit parameters in sampling from Eqs.~\ref{eqn:mclmgal}-\ref{eqn:rhmcl}\footnote{We assume an average compactness parameter $r_{\rm h}/r_{\rm c}=10$ \citep{GeorgievBoker2014}.}. From the NSC mass and size, we compute the escape velocity from the center \citep[e.g.,][]{FragioneSilk2020,FragioneKocsis2022}
\begin{equation}
\vesc \approx 50\kms\left(\frac{M_{\rm CL}}{10^5\msun}\right)^{1/2}\left(\frac{r_h}{1\,{\rm pc}}\right)^{-1/2}\,,
\label{eqn:mvesc}
\end{equation}

Hereafter, we summarize our numerical procedure to follow the growth and mergers of IMBHs, starting from a BH seed of mass $\mseed$, which undergoes mergers with other stellar-mass BHs \citep[for details see][]{FragioneKocsis2022}.

\begin{enumerate}

    \item \textit{Initial seed mass}. We either fix the mass of the growing BH seed to some initial value \citep[e.g., produced through repeated mergers of massive stars][]{GonzalezKremer2021,WeatherfordFragione2021} or we derive it directly from stellar evolution. In the latter case, we sample the stellar mass of the seed progenitor from a \citet{kro01} initial mass function\footnote{We assume the progenitor seed is born with no companion. \citet{FragioneKocsis2022} showed that changing the primordial binary fractions of stars that are BH progenitors does not have a significant impact on the mass distribution or on the relative outcomes of IMBHs.} and evolve it using \textsc{sse} \citep{HurleyPols2000,BanerjeeBelczynski2020}. If the seed is not ejected by natal kicks, $v_{\rm natal}$, as a result of asymmetric supernova explosion, that is $v_{\rm natal}<\vesc$, we compute the timescale for the seed to sink to the cluster center via dynamical friction \citep{Chandrasekhar1943}.
    
    \item \textit{Formation of binaries and mergers}. We estimate the timescale for the seed to find a BH companion by accounting for three-body binary formation and binary-single encounters, whichever is the faster depending on the NSC mass and density \citep[e.g.,][]{antoras2016}. We draw the mass of the companion, $m_2$, by considering that the pairing probability scales as $\mathcal{P}\propto (\mseed+m_2)^4$ \citep[e.g.,][]{omk16}. After the binary is formed, we compute the timescale for it to shrink via three-body interactions until it reaches a critical semi-major axis, $\max(a_{\rm ej},a_{\rm GW})$ \citep[see Eqs.~21-22 in][]{FragioneKocsis2022}. If $a_{\rm ej}>a_{\rm GW}$, the binary is ejected from the cluster, halting further mergers; if $a_{\rm ej}<a_{\rm GW}$, we compute the timescale to merge via GW emission.
    
    \item \textit{Retention and repeated mergers}. We compute the remnant spin and mass, and the relativistic recoil kick imparted to it as a result of asymmetry in GW emission \citep[e.g.,][]{lou08,lou10,lou12,HofmannBarausse2016,Jimenez-FortezaKeitel2017}. If the binary is retained within the cluster, that is the recoil kick, $v_{\rm kick}$, is smaller than the NSC escape speed ($v_{\rm kick}<\vesc$), we compute the timescales for the seed to sink back to the cluster center via dynamical friction \citep{Chandrasekhar1943}. If the total elapsed time is smaller than $10$ Gyr, we repeat from step (2) and generate hierarchical mergers.
    
\end{enumerate}
For details about the various parameters and important quantities adopted in our model, see Table~1 in \citet{FragioneKocsis2022}.

We run three different models corresponding to different assumptions for the initial mass of the BH seed. In the first model, we randomly sample the initial seed mass from a realistic mass spectrum for stellar BH remnants. In this case, the average initial seed mass would be about $10\msun$, with little dependence on the cluster metallicity. In the second and third models, we assume an initial seed mass of $50\msun$ and  $100\msun$, respectively. The latter two models could represent the case where a massive seed is produced in the beginning of the cluster lifetime as a result of the collapse of a very massive star, formed through repeated stellar mergers \citep[e.g.,][]{por02,GonzalezKremer2021}. The birth spin of BHs is quite uncertain. To encompass all the possibilities, we adopt four different fixed values for the initial spin parameter of BHs, namely $\chi=0$, $0.2$, $0.5$, $0.8$. For each model, we consider eight different metallicities ($Z=0.0001$, $0.0002$, $0.0005$, $0.001$, $0.002$, $0.005$, $0.01$, and $0.02$). For each combination, we run $50$k simulations, for a total of $4.8\times 10^7$.

\section{Results}
\label{sect:results}

\subsection{Merger rates}

\begin{figure} 
\centering
\includegraphics[scale=0.59]{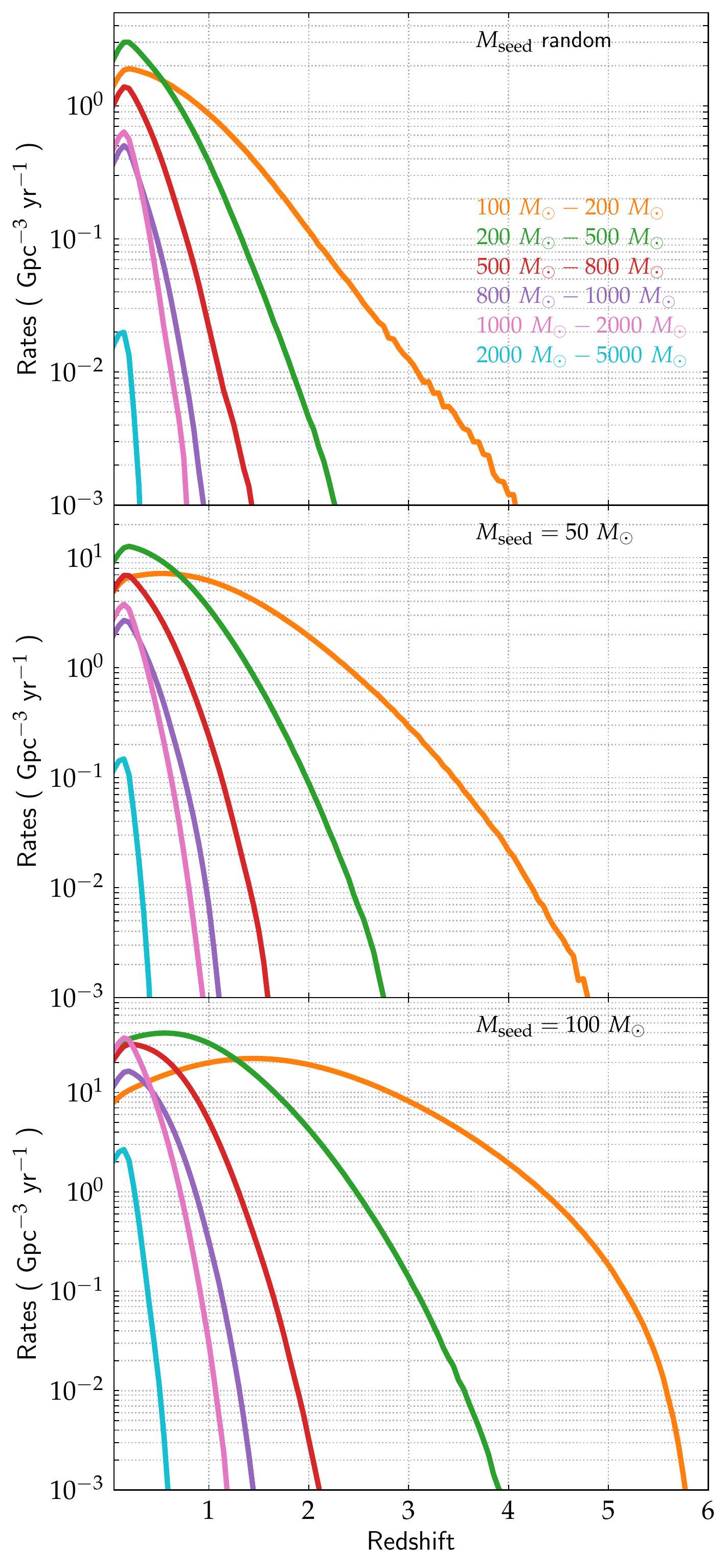}
\caption{Volumetric merger rate for IMBH binaries for different initial choices of the seed mass: random initial seed mass in the range of stellar-mass BHs (top), initial seed mass of $50\msun$ (center), and initial seed mass of $100\msun$ (bottom). The initial BH spins are assumed to be zero. Different colors represent different IMBH masses.}
\label{fig:mseed}
\end{figure}

We compute the differential merger rate of IMBH binaries as follows. We start with computing the merger rate as
\begin{eqnarray}
R(z) &=& \frac{d}{dt_{\rm lb}(z)} \int^{z_{\max}}_{z} \Psi_{\rm NSC}(\zeta) \frac{dt_{\rm lb}(\zeta)}{d\zeta} d\zeta \nonumber\\
& \times & \int^{Z_{\max}(\zeta)}_{Z_{\min}(\zeta)} \Phi(\zeta, Z) \Pi(\zeta, Z) dZ\,,
\label{eqn:ratez}
\end{eqnarray}
where $z_{\max}=6$, $t_{\rm lb}$ is the look-back time at redshift $z$ \footnote{For our calculations we assume the cosmological parameters from Planck 2015 \citep{PlanckCollaborationAde2016}.}, $\Psi_{\rm NSC}(z)$ is the cosmic NSC formation history, $\Phi(z, Z)$ is the merger efficiency at a given metallicity $Z$, and $\Pi(z, Z)$ is the metallicity distribution at a given redshift, which we assume is described by a log-normal distribution with mean given by \citep{MadauFragos2017}
\begin{equation}
\log \langle Z/{\rm Z}_\odot \rangle = 0.153 - 0.074 z^{1.34}
\end{equation}
and a standard deviation of $0.5$~dex \citep{DvorkinSilk2015}. The formation history of NSCs is highly uncertain \citep[e.g.,][]{NeumayerSeth2020}. Here, we model it as
\begin{equation}
    \Psi_{\rm NSC}(z) = \mathcal{B}_{\rm NSC} \exp\left[-(z-z_{\rm NSC})^2 / (2\sigma_{\rm NSC}^2) \right]\,.
    \label{eqn:sf}
\end{equation}
In the previous equation, we assume $\mathcal{B}_{\rm NSC}= 10^{-5}\msun\,{\rm Mpc}^{-3}\,{\rm yr}^{-1}$ \citep{MapelliDall'Amico2021}, chosen so that the NSC density in the local Universe is consistent with the observed one. We also fix $z_{\rm NSC}=3.2$ and $\sigma_{\rm NSC}=1.5$, respectively, following the cosmic formation history of globular clusters \citep[e.g.,][]{GrattonFusiPecci1997,GrattonBragaglia2003,VandenBergBrogaard2013,El-BadryQuataert2019}, under the assumption that NSCs form mostly from the inspiral of globular clusters into the galactic through dynamical friction \citep[e.g.,][]{capuzz2008,ant2014}. Finally, we model the merger efficiency at a given metallicity $Z$ as \citep[e.g.,][]{MapelliDall'Amico2021}
\begin{equation}
\Phi(z, Z) = \epsilon(z, Z)\frac{N_{\rm BH} (Z)}{M_{\rm tot}(Z)} \,,
\end{equation}
where $\epsilon(z, Z)$ is the number of mergers of IMBH binaries per simulated system at a given redshift, $N_{\rm BH} (Z)$ the total number of BH at a given metallicity, assuming a \citet{kro01} initial mass function, and $M_{\rm tot} (Z)$ is the total simulated NSC mass at metallicity $Z$. We compute Eq.~\ref{eqn:ratez} for different IMBH mass bins, that is $R_{i}(z)$. 

Figure~\ref{fig:mseed} shows the volumetric merger rate for IMBH binaries for different initial choices of the seed mass, assuming that the initial BH spins are zero. In the first model, we find that IMBH merger rates are peaked at $z\lesssim 1$, and are about $1.5\gpcyr$, $1\gpcyr$, $2\gpcyr$, $0.4\gpcyr$, $0.5\gpcyr$ for IMBH in the mass bins $100\msun-200\msun$, $200\msun-500\msun$, $500\msun-800\msun$, $800\msun-1000\msun$, $1000\msun-2000\msun$, $2000\msun-5000\msun$, respectively, in the local Universe. The assumption on the initial seed mass affects the overall merger rate and the relative rates of different mass bins. For $\mseed=50\msun$ and $100\msun$, the overall merger rate increases by a factor of about $5$ and $25$, respectively, and the merger rate of IMBHs with masses $\gtrsim 500\msun$ becomes comparable to the merger rate of IMBHs with masses $\lesssim 500\msun$. This trend can be explained considering that the recoil kick imparted to the growing seed becomes smaller for larger initial seed masses, therefore rendering less likely the ejection of the growing IMBH. Our estimates are consistent (within current uncertainties) with LIGO/Virgo/KAGRA upper limits for mergers of GW190521-like systems \citep{ligo2020new2} and IMBH binaries with masses up to about $500\msun$ \citep{AbbottAbbott2017imbh,AbbottAbbott2019,AbbottAbbott2022imbh}.

\begin{figure} 
\centering
\includegraphics[scale=0.59]{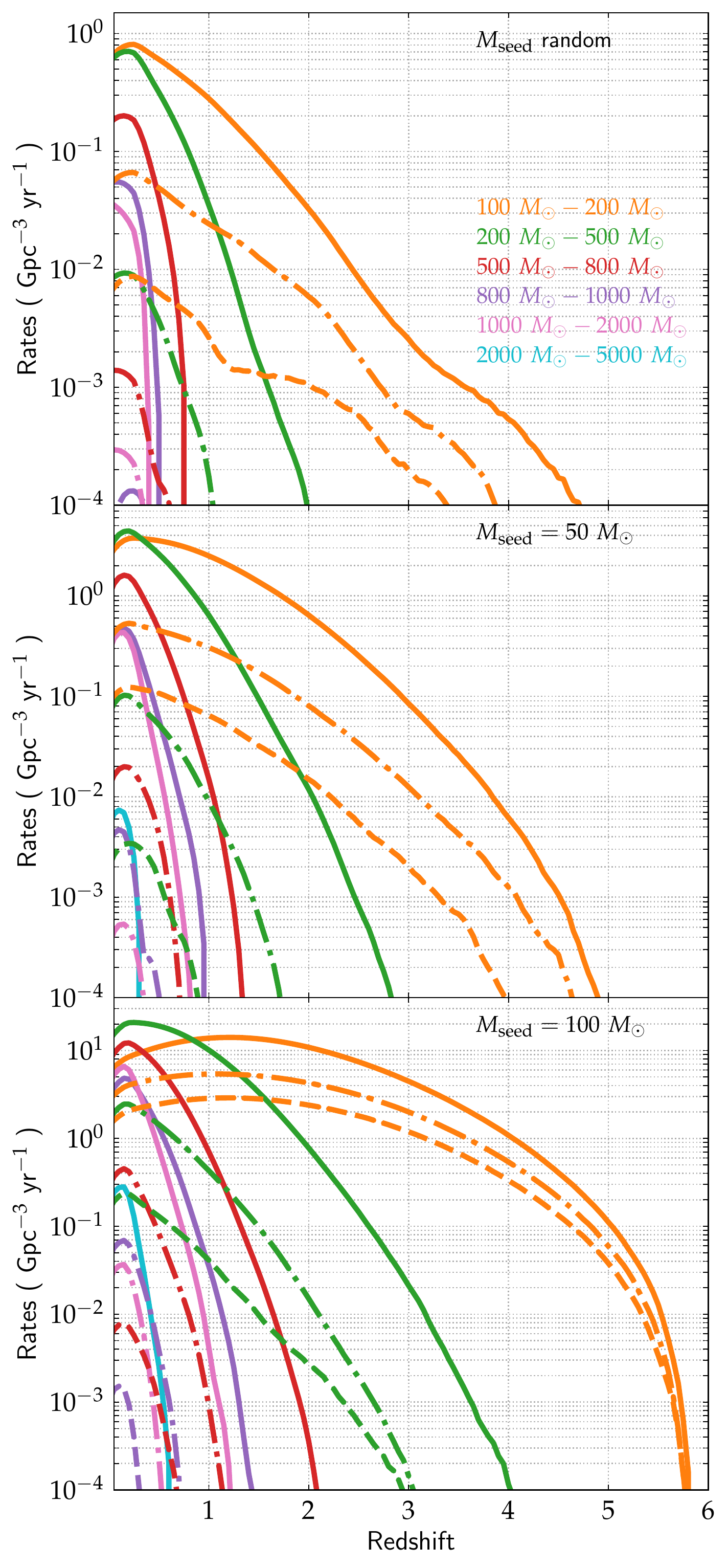}
\caption{Same as Fig.~1 but here the initial BH spin parameter is assumed to be $\chi=0.2$ (solid lines), $0.5$ (dot-dashed lines), or $0.8$ (dashed lines).}
\label{fig:spin}
\end{figure}

We show the effect of the initial BH spin on the expected merger rate in Figure~\ref{fig:spin}. Since recoil kicks critically depend on BH spins (as long as the mass ratio of a merger is $\gtrsim 0.1$), the rates are significantly affected by spins. If BHs are born with non-zero spins, the growing seed can be more efficiently ejected from the host star cluster, halting further growth. Since recoil kicks also depend on the mass ratio of the merging binary, merger rates are less affected by a non-zero initial BH spin for larger initial seed masses. For example, for masses in the range $100\msun-200\msun$, the merger rates become almost two orders of magnitude smaller when the initial BH spin parameter goes from $0.2$ to $0.8$ for an initial random seed, while they go down by a much smaller factor of about $4$ when the initial seed mass is $100\msun$. Finally, the effect of the initial BH spin is more important on the merger rates of larger IMBHs. Indeed, larger initial spins imply larger recoil kicks, which can eject the growing seed from the parent cluster, reducing the merger rates of large IMBHs.

\subsection{Dependence on NSC formation history}

The formation history of NSCs is highly uncertain. As shown in Eq.~\ref{eqn:sf}, we model it as a Gaussian distribution with mean $z_{\rm NSC}=3.2$ and variance $\sigma_{\rm NSC}=1.5$. To analyze how our results depend on these assumptions, we run two additional models where we set $z_{\rm NSC}=1.6$ ($\mathcal{B}_{\rm NSC}= 4.9\times 10^{-4}\msun\,{\rm Mpc}^{-3}\,{\rm yr}^{-1}$) and $z_{\rm NSC}=5.2$ ($\mathcal{B}_{\rm NSC}= 3.7\times 10^{-5}\msun\,{\rm Mpc}^{-3}\,{\rm yr}^{-1}$), respectively. These two additional models represent a later and earlier NSC formation peak, respectively, with respect to our main model.

\begin{figure} 
\centering
\includegraphics[scale=0.59]{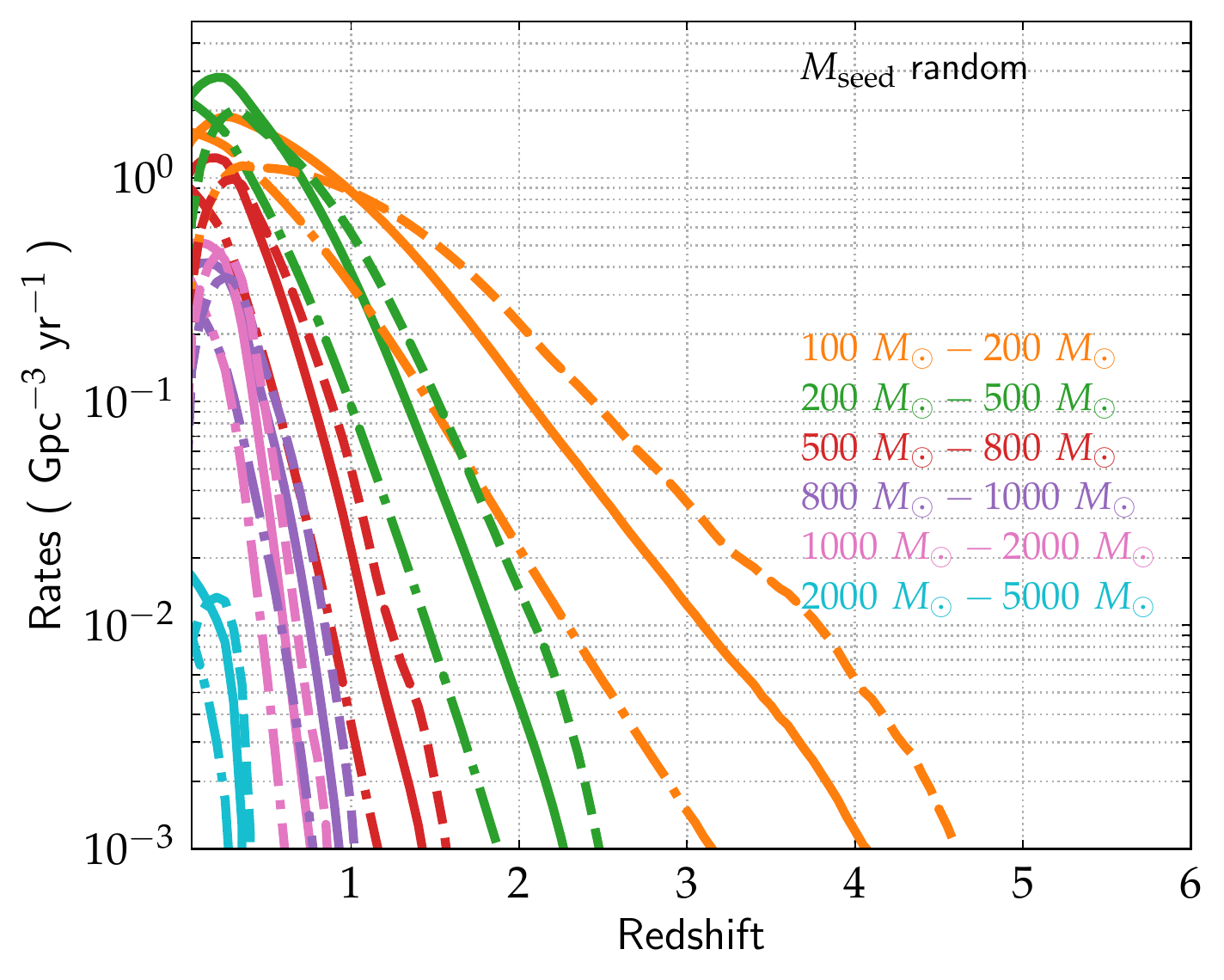}
\caption{Volumetric merger rate for IMBH binaries for random initial seed mass in the mass spectrum of stellar-mass BHs and different NSC formation histories ($\sigma_{\rm NSC}=1.5$; see Eq.~\ref{eqn:sf}): $z_{\rm NSC}=3.2$ (solid), $z_{\rm NSC}=1.6$ (dot-dashed), $z_{\rm NSC}=5.2$ (dashed). The initial BH spins are assumed to be zero. Different colors represent different IMBH masses.}
\label{fig:sf}
\end{figure}

We show the results of our additional runs in Figure~\ref{fig:sf}, in the case of a random initial seed mass in the mass spectrum of stellar-mass BHs. We find that our various formation histories do not have a significant impact on the magnitude of the IMBH merger rate, while affecting its shape for masses $\lesssim 500\msun$.

\begin{figure} 
\centering
\includegraphics[scale=0.59]{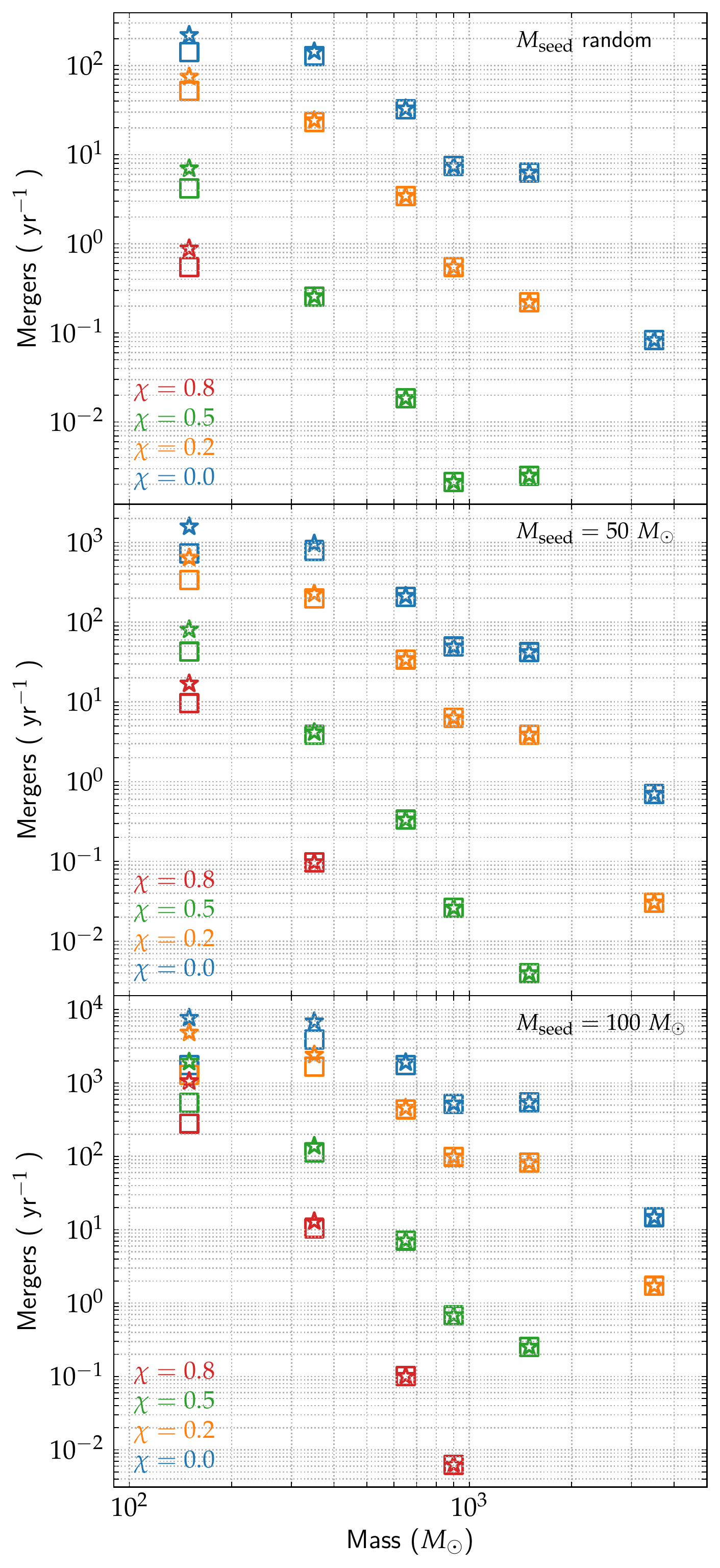}
\caption{Cumulative merger rate as a function of the IMBH mass for different choices of the initial seed mass: random initial seed mass in the mass spectrum of stellar-mass BHs (top), initial seed mass of $50\msun$ (center), and initial seed mass of $100\msun$ (bottom). Different colors represent different initial BH spins. Squares: $z=1$; stars: $z=3$.}
\label{fig:mergers}
\end{figure}

\subsection{Number of mergers}

The number of mergers we observed per year is given by the cumulative merger rate
\begin{equation}
    C(z)=\int_0^z R(\zeta) \frac{dV_{c}(\zeta)}{d\zeta}(1+\zeta)^{-1}d\zeta\,,
\end{equation}
where $(dV_{c}(z)/dz)$ is the amount of comoving volume in a slice of the Universe at redshift $z$ and $(1+z)^{-1}$ is the difference in comoving time between the merger redshift and the observer at $z=0$. 

In Figure~\ref{fig:mergers}, we show the cumulative merger rate as a function of the IMBH mass for different choices of the initial seed mass and initial BH spins, at redshift $z=1$ (squares) and $z=3$ (stars). As expected from the analysis of Figures~\ref{fig:mseed}-\ref{fig:spin}, most of the binary mergers take place at low redshift, $z\lesssim 1$, with the exception of the smallest IMBHs with masses $\lesssim 300\msun$. In this case, the number of events is higher by a factor of a few at $z=3$ with respect to $z=1$, with the largest differences attained for larger initial seed masses and smaller initial BH spins. Mergers of IMBHs with masses $\gtrsim 300\msun$ occur at $z<1$, as a result of the progressive building up of massive IMBHs.

\begin{figure*} 
\centering
\includegraphics[scale=0.41]{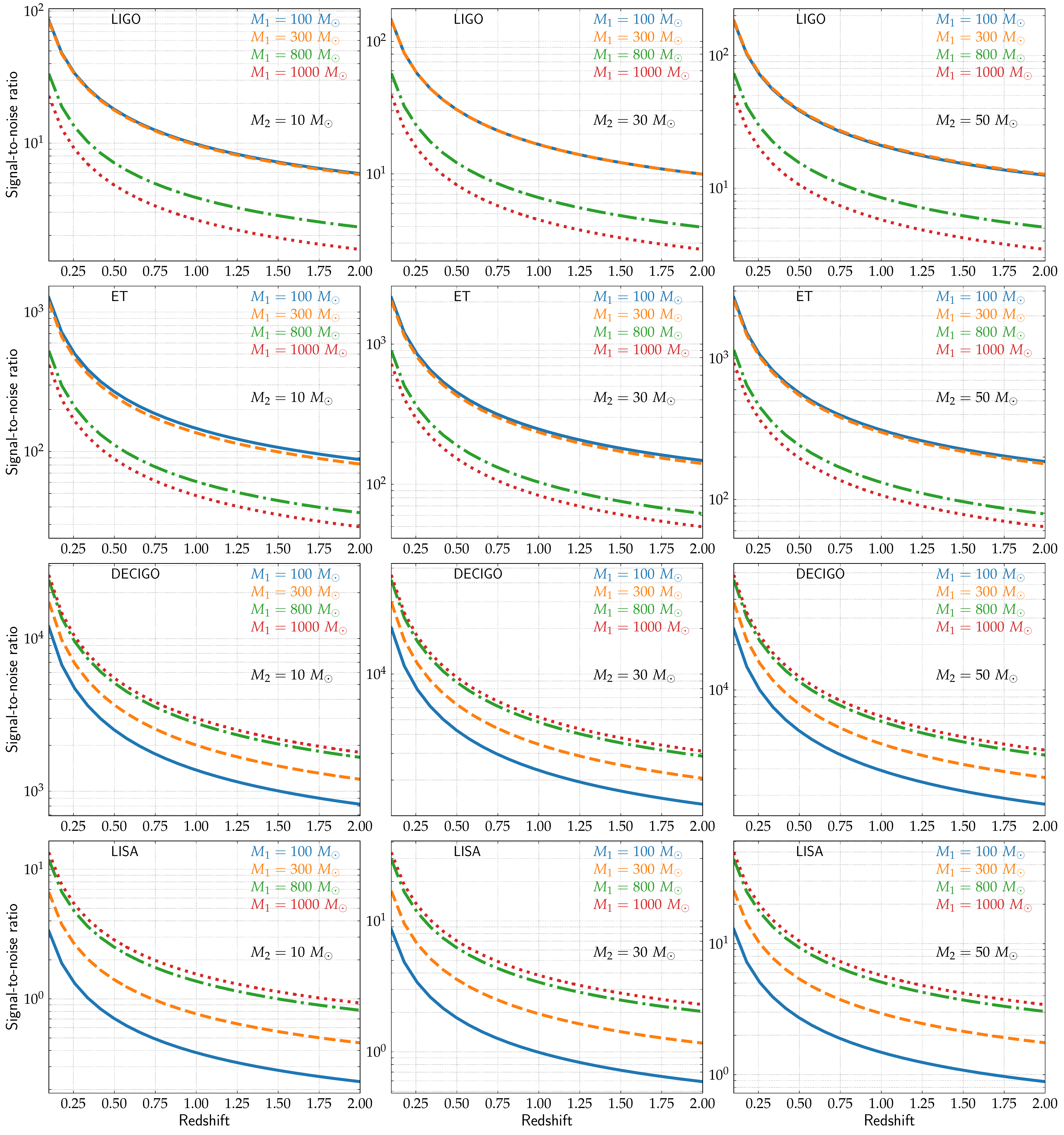}
\caption{Average signal-to-noise ratio of IMBH binaries as a function of redshift for different detectors: LIGO (top), ET (center-top), DECIGO (center-bottom), LISA (bottom). Different colors represent different IMBH masses, while different columns represent different secondary masses.}
\label{fig:snr}
\end{figure*}

\begin{figure*} 
\centering
\includegraphics[scale=0.475]{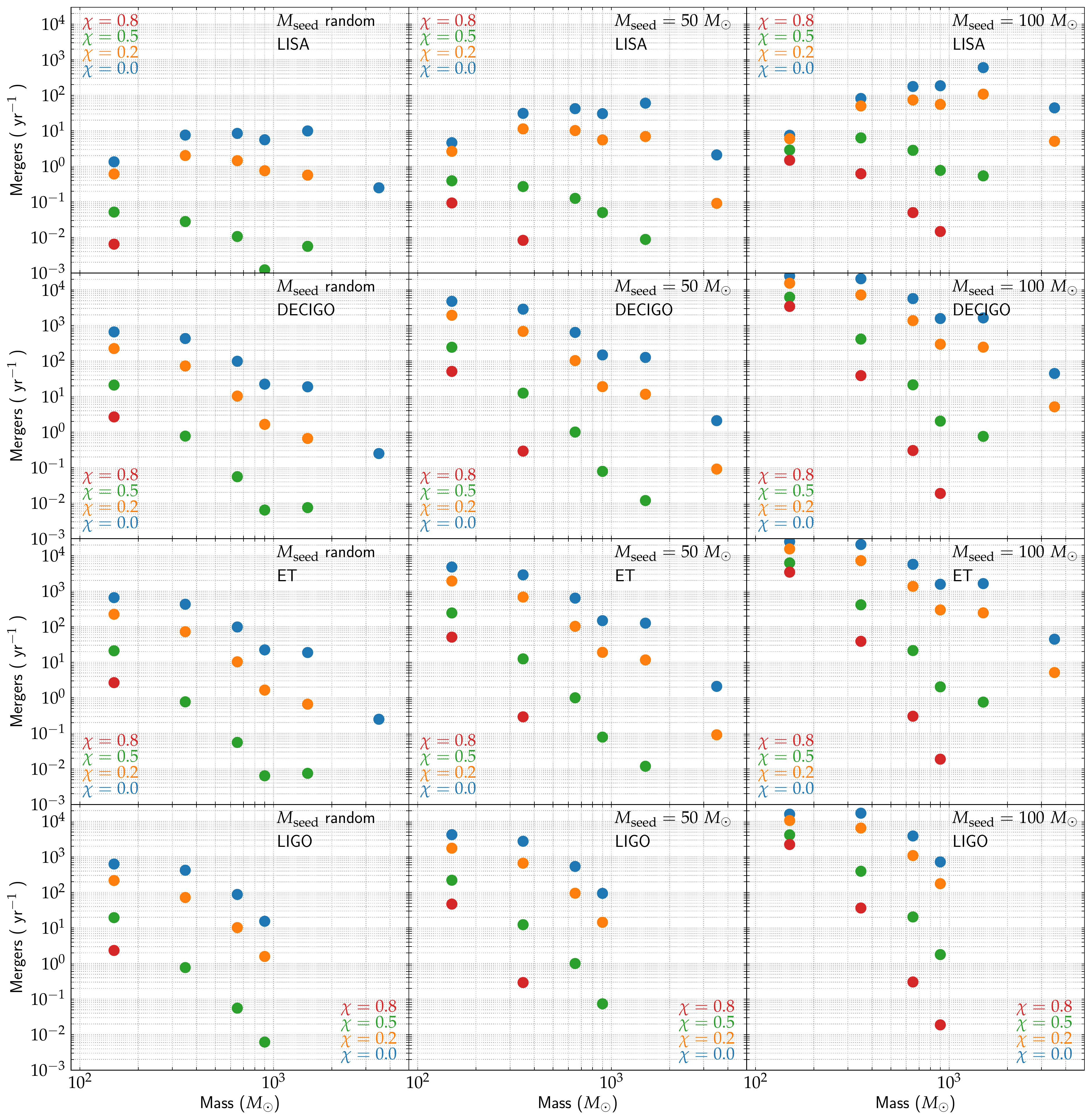}
\caption{Maximum number of detected merger events per year as a function of the IMBH mass for different GW observatories: random initial seed mass in the mass spectrum of stellar-mass BHs (left), initial seed mass of $50\msun$ (center), and initial seed mass of $100\msun$ (right). Different colors represent different initial BH spins.}
\label{fig:mergers_detect}
\end{figure*}

\subsection{Multi-band detections}

We compute the signal-to-noise ratio (SNR) of our merging systems in a detector frequency band, to understand the prospects for multi-band detections by ground-based and space-based observatories. Given the masses of the merging BHs, $m_1$ and $m_2$, we compute the average SNR as
\begin{equation}
\left\langle \frac{S}{N} \right\rangle=\frac{4}{\sqrt{5}} \sqrt{\int_{f_{\rm min}}^{f_{\rm max}} \frac{|\tilde{h}(f)|^2}{S_n(f)} df}\,,
\label{eqn:rhof}
\end{equation}
where $f_{\rm min}$ and $f_{\rm max}$ are the minimum and maximum frequency of the binary in the dectector band, respectively, $S_n(f)$ is the effective noise power spectral density, and $|\tilde{h}(f)|$ is the frequency-domain waveform amplitude for a face-on binary, approximated with a PhenomA waveform \citep[e.g., Eq.~20 in][]{robson2019}
\begin{eqnarray}
|\tilde{h}(f)| &=& \sqrt{\frac{5}{24\pi^{4/3}}} \frac{G^{5/6}}{c^{3/2}} \frac{M_{\rm c,z}^{5/6}}{D_{\rm L}f_0^{7/6}}\nonumber\\
&\times&
\begin{dcases}
(f/f_0)^{-7/6} & f<f_0\\
(f/f_0)^{-2/3} & f_0\leq f<f_1\\
w\mathcal{L}(f,f_1,f_2) & f_1\leq f<f_3\,.
\end{dcases}
\label{eqn:waveform}
\end{eqnarray}
In the previous equation,
\begin{eqnarray}
f_k &= & \frac{a_k\eta^2+b_k\eta+c_k}{\pi(GM_z/c^3)},\\
\mathcal{L} &=& \left(\frac{1}{2\pi}\right) \frac{f_2}{(f-f_1)^2+f_2^2/4},\\
w &= & \frac{\pi f_2}{2} \left(\frac{f_0}{f_1}\right)^{2/3}\,,
\end{eqnarray}
where the values of $\{f_k,a_k,b_k,c_k\}$ are taken from Table 2 in \citet{robson2019}, $f$ is the detector-frame frequency, related to the binary orbital frequency by $f=f_{\rm orb}(1+z)^{-1}$, $M_{\rm c,z}$ is the redshifted chirp mass, related to the rest-frame chirp mass
\begin{equation}
M_\mathrm{c}=\frac{m_1^{3/5}m_2^{3/5}}{(m_1+m_2)^{1/5}}
\end{equation}
by $M_\mathrm{c}=M_\mathrm{c,z}/(1+z)$, and
\begin{equation}
D_{\rm L}=(1+z)\frac{c}{H_0}\int_{0}^z \frac{d\zeta}{\sqrt{\Omega_{\rm M}(1+\zeta^3)+\Omega_\Lambda}}
\end{equation}
is the luminosity distance, with $c$ and $H_0$ being the velocity of light and Hubble constant, respectively, and $\Omega_{\rm M}=0.286$ and $\Omega_\Lambda=0.714$ \citep{PlanckCollaborationAde2016}), respectively. We compute the power spectral density of LISA as in \citet{robson2019}, of ET as in \citep{HildAbernathy2011}, of DECIGO as in \citep{YagiSeto2011}, and of LIGO at design sensitivity as in\citep{Ajith2011}. In order to asses the possibility of multi-band detections, we choose $f_{\rm min}=\max(f_{\rm mb},f_{\rm min, j})$, where $f_{\rm min, j}$ is the minimum frequency in a detector band ($10^{-5}$\,Hz, $10^{-3}$\,Hz, $1$\,Hz, $10$\,Hz for LISA, DECIGO, ET, LIGO, respectively) and 
\begin{eqnarray}
    f_{\rm mb} &=& 0.04\,{\rm Hz} \left(\frac{m_1 + m_2}{100\msun}\right)^{1/8}\nonumber\\
    & \times &\left(\frac{m_1 m_2}{100\msun}\right)^{-3/8} \left(\frac{T_{\rm LISA}}{4\,{\rm yr}}\right)^{-3/8}
\end{eqnarray}
is the minimum initial frequency for a binary to merge within the LISA mission lifetime, which we set to $T_{\rm LISA}=4$\,yr. We set as a minimum detection threshold $\langle S/N \rangle_{\rm thr}=8$.

We show in Figure~\ref{fig:snr} the expected average SNR of IMBH binaries as a function of redshift for different detectors and various masses of the IMBH and its companion. The average SNR in both LIGO and ET is generally higher for larger companion masses and smaller IMBH masses, as expected since binaries with more massive IMBHs will spend fewer cycles in band. With its smaller characteristic noise, ET will be able to detect these binaries further than LIGO. For instance, a merging IMBH binary is expected to be detected with an SNR $\gtrsim 10$ by ET even at $z=2$, while the corresponding SNR in LIGO is $\lesssim 10$. On the contrary, DECIGO and LISA will be able to detect merging binaries to larger distances for larger IMBH and secondary masses. While DECIGO is expected to detect such systems with SNR $\gtrsim 10^3$ even at redshift $z=2$, LISA will essentially be able to detect merging binaries with IMBH mass $\gtrsim 800\msun$ and companion masses $\gtrsim 30\msun$ at $z\lesssim 0.5$.\footnote{Note that LISA could observe binaries containing an IMBH earlier during the inspiral to larger distances (not shown). Also note that the SNR is higher by a factor $2.5\times$ for an optimal orientation of the binary relative to the average shown in the figure.}

We show in Figure~\ref{fig:mergers_detect} the maximum number of detected merger events per year as a function of the IMBH mass for different GW observatories, for various choices of the initial seed mass and BH spin. As discussed, most of the binary mergers take place at low redshift, $z\lesssim 1$, with the exception of the smallest IMBHs with masses $\lesssim 300\msun$, for which a nice fraction of events comes from larger redshifts. Our model predicts that a few merger events per year should be detectable with LISA, DECIGO, ET, and LIGO for IMBHs with masses $\lesssim 1000\msun$, and a few tens of merger events per year with DECIGO, ET, and LIGO only. 

\section{Discussion and conclusions}
\label{sect:concl}

GW detectors are expected to revolutionize our understanding of the elusive IMBHs, after establishing firmly their existence.

We have analyzed the statistical results of a wide range of simulations, which encompass different metallicities, initial seed masses, initial BH spins, run with the semi-analytical code developed in \citet{FragioneKocsis2022}. We have computed the merger rate of IMBH binaries and have found that rates are in the range $0.01$\,Gpc$^{-3}$\,yr$^{-1}$-$10$\,Gpc$^{-3}$\,yr$^{-1}$ for different IMBH masses. We have also computed the number of multi-band detections in ground-based and space-based observatories. Our model predicts that a few merger events per year should be detectable with LISA, DECIGO, ET, and LIGO for IMBHs with masses $\lesssim 1000\msun$, and a few tens of merger events per year with DECIGO, ET, and LIGO only. Our estimates are consistent (within current uncertainties) with LIGO/Virgo/KAGRA upper limits for mergers of GW190521-like systems \citep{ligo2020new2} and IMBH binaries with masses up to about $500\msun$ \citep{AbbottAbbott2017imbh,AbbottAbbott2019,AbbottAbbott2022imbh}. As the sensitivity of current detectors is improved and new observatories start operating, our models predict tens or hundreds of merging binary signals in the next few years. In the case of fewer detections, the inferred observational merger rates can be used to constrain the model parameters of NSCs, their formation histories, and IMBH spins.

As already widely discussed in \citet{FragioneKocsis2022}, there are some limitations to our approach. For example, we have assumed that the masses and sizes of NSCs do not evolve significantly during their lifetime and that NSCs are isolated with respect to the rest of their host galaxy history. In reality, NSCs are not isolated from their environments and evolve on a cosmological timescale, as there can be continuous supply of stars and gas from the rest of the galaxy, with ongoing star formation and accretion of smaller star clusters. The NSC evolution could be significantly affected by the presence of a (single or binary) SMBH and/or gas \citep[e.g.,][]{bahc1976,hopale2006,ant2014,VogelsbergerGenel2014,alex2017}. Despite these limitations, we believe that our model gives reasonable order-of-magnitude estimates for IMBH mergers in NSCs. 

\section*{Acknowledgements}

G.F.\ and F.A.R.\ acknowledge support from NASA Grant 80NSSC21K1722. A.L.\ was supported in part by the Black Hole Initiative at Harvard University, which is funded by JTF and GBMF grants. This work also received funding from the European Research Council (ERC) under the European Union’s Horizon 2020 Programme for Research and Innovation ERC-2014-STG under grant agreement No.~638435 (GalNUC) (to B.K.).

\bibliographystyle{aasjournal}
\bibliography{refs}

\begin{thebibliography}{}
\expandafter\ifx\csname natexlab\endcsname\relax\def\natexlab#1{#1}\fi
\providecommand{\url}[1]{\href{#1}{#1}}
\providecommand{\dodoi}[1]{doi:~\href{http://doi.org/#1}{\nolinkurl{#1}}}
\providecommand{\doeprint}[1]{\href{http://ascl.net/#1}{\nolinkurl{http://ascl.net/#1}}}
\providecommand{\doarXiv}[1]{\href{https://arxiv.org/abs/#1}{\nolinkurl{https://arxiv.org/abs/#1}}}

\bibitem[{{Abbott} {et~al.}(2019){Abbott}, {Abbott}, {Abbott}, {Abraham}, \&
  et~al.}]{AbbottAbbott2019}
{Abbott}, B.~P., {Abbott}, R., {Abbott}, T.~D., {Abraham}, S., \& et~al. 2019,
  \prd, 100, 064064, \dodoi{10.1103/PhysRevD.100.064064}

\bibitem[{{Abbott} {et~al.}(2017){Abbott}, {Abbott}, {Abbott}, {Acernese}, \&
  et~al.}]{AbbottAbbott2017imbh}
{Abbott}, B.~P., {Abbott}, R., {Abbott}, T.~D., {Acernese}, F., \& et~al. 2017,
  \prd, 96, 022001, \dodoi{10.1103/PhysRevD.96.022001}

\bibitem[{{Abbott} {et~al.}(2020){Abbott}, {Abbott}, {Abraham}, {Acernese}, \&
  et~al.}]{AbbottAbbott1905212020}
{Abbott}, R., {Abbott}, T.~D., {Abraham}, S., {Acernese}, F., \& et~al. 2020,
  \prl, 125, 101102, \dodoi{10.1103/PhysRevLett.125.101102}

\bibitem[{{Abbott} {et~al.}(2022){Abbott}, {Abbott}, {Acernese}, {Ackley}, \&
  et~al.}]{AbbottAbbott2022imbh}
{Abbott}, R., {Abbott}, T.~D., {Acernese}, F., {Ackley}, K., \& et~al. 2022,
  \aap, 659, A84, \dodoi{10.1051/0004-6361/202141452}

\bibitem[{{Ajith}(2011)}]{Ajith2011}
{Ajith}, P. 2011, \prd, 84, 084037, \dodoi{10.1103/PhysRevD.84.084037}

\bibitem[{{Alexander}(2017)}]{alex2017}
{Alexander}, T. 2017, \araa, 55, 17,
  \dodoi{10.1146/annurev-astro-091916-055306}

\bibitem[{{Amaro-Seoane} {et~al.}(2017){Amaro-Seoane}, {Audley}, {Babak},
  {Baker}, \& et~al.}]{Amaro-SeoaneAudley2017}
{Amaro-Seoane}, P., {Audley}, H., {Babak}, S., {Baker}, J., \& et~al. 2017,
  arXiv e-prints, arXiv:1702.00786.
\newblock \doarXiv{1702.00786}

\bibitem[{{Antonini}(2014)}]{ant2014}
{Antonini}, F. 2014, \apj, 794, 106, \dodoi{10.1088/0004-637X/794/2/106}

\bibitem[{{Antonini} {et~al.}(2019){Antonini}, {Gieles}, \&
  {Gualandris}}]{antonini2019}
{Antonini}, F., {Gieles}, M., \& {Gualandris}, A. 2019, \mnras, 486, 5008,
  \dodoi{10.1093/mnras/stz1149}

\bibitem[{{Antonini} \& {Rasio}(2016)}]{antoras2016}
{Antonini}, F., \& {Rasio}, F.~A. 2016, \apj, 831, 187,
  \dodoi{10.3847/0004-637X/831/2/187}

\bibitem[{{Arca Sedda} {et~al.}(2021){Arca Sedda}, {Amaro Seoane}, \&
  {Chen}}]{ArcaSeddaAmaroSeoane2021}
{Arca Sedda}, M., {Amaro Seoane}, P., \& {Chen}, X. 2021, \aap, 652, A54,
  \dodoi{10.1051/0004-6361/202037785}

\bibitem[{{Bahcall} \& {Wolf}(1976)}]{bahc1976}
{Bahcall}, J.~N., \& {Wolf}, R.~A. 1976, \apj, 209, 214, \dodoi{10.1086/154711}

\bibitem[{{Baldassare} {et~al.}(2015){Baldassare}, {Reines}, {Gallo}, \&
  {Greene}}]{BaldassareReines2015}
{Baldassare}, V.~F., {Reines}, A.~E., {Gallo}, E., \& {Greene}, J.~E. 2015,
  \apjl, 809, L14, \dodoi{10.1088/2041-8205/809/1/L14}

\bibitem[{{Banerjee} {et~al.}(2020){Banerjee}, {Belczynski}, {Fryer},
  {Berczik}, \& et~al.}]{BanerjeeBelczynski2020}
{Banerjee}, S., {Belczynski}, K., {Fryer}, C.~L., {Berczik}, P., \& et~al.
  2020, \aap, 639, A41, \dodoi{10.1051/0004-6361/201935332}

\bibitem[{{Begelman} {et~al.}(2006){Begelman}, {Volonteri}, \&
  {Rees}}]{begelm2006}
{Begelman}, M.~C., {Volonteri}, M., \& {Rees}, M.~J. 2006, \mnras, 370, 289,
  \dodoi{10.1111/j.1365-2966.2006.10467.x}

\bibitem[{{Bromm} \& {Larson}(2004)}]{bromm2004}
{Bromm}, V., \& {Larson}, R.~B. 2004, \araa, 42, 79,
  \dodoi{10.1146/annurev.astro.42.053102.134034}

\bibitem[{{Bromm} \& {Loeb}(2003)}]{bromm2003}
{Bromm}, V., \& {Loeb}, A. 2003, \apj, 596, 34, \dodoi{10.1086/377529}

\bibitem[{{Capuzzo-Dolcetta} \& {Miocchi}(2008)}]{capuzz2008}
{Capuzzo-Dolcetta}, R., \& {Miocchi}, P. 2008, \mnras, 388, L69,
  \dodoi{10.1111/j.1745-3933.2008.00501.x}

\bibitem[{{Chandrasekhar}(1943)}]{Chandrasekhar1943}
{Chandrasekhar}, S. 1943, \apj, 97, 255, \dodoi{10.1086/144517}

\bibitem[{{Chen} {et~al.}(2011){Chen}, {Sesana}, {Madau}, \& {Liu}}]{chen2011}
{Chen}, X., {Sesana}, A., {Madau}, P., \& {Liu}, F.~K. 2011, \apj, 729, 13,
  \dodoi{10.1088/0004-637X/729/1/13}

\bibitem[{{Chilingarian} {et~al.}(2018){Chilingarian}, {Katkov}, {Zolotukhin},
  {Grishin}, {Beletsky}, {Boutsia}, \& {Osip}}]{chili2018}
{Chilingarian}, I.~V., {Katkov}, I.~Y., {Zolotukhin}, I.~Y., {et~al.} 2018,
  \apj, 863, 1, \dodoi{10.3847/1538-4357/aad184}

\bibitem[{{Das} {et~al.}(2021){Das}, {Schleicher}, {Leigh}, \&
  {Boekholt}}]{DasSchleicher2021}
{Das}, A., {Schleicher}, D. R.~G., {Leigh}, N. W.~C., \& {Boekholt}, T. C.~N.
  2021, \mnras, 503, 1051, \dodoi{10.1093/mnras/stab402}

\bibitem[{{Di Carlo} {et~al.}(2021){Di Carlo}, {Mapelli}, {Pasquato},
  {Rastello}, \& et~al.}]{DiCarloMapelli2021}
{Di Carlo}, U.~N., {Mapelli}, M., {Pasquato}, M., {Rastello}, S., \& et~al.
  2021, \mnras, \dodoi{10.1093/mnras/stab2390}

\bibitem[{{Dvorkin} {et~al.}(2015){Dvorkin}, {Silk}, {Vangioni}, {Petitjean},
  \& et~al.}]{DvorkinSilk2015}
{Dvorkin}, I., {Silk}, J., {Vangioni}, E., {Petitjean}, P., \& et~al. 2015,
  \mnras, 452, L36, \dodoi{10.1093/mnrasl/slv085}

\bibitem[{{El-Badry} {et~al.}(2019){El-Badry}, {Quataert}, {Weisz}, {Choksi},
  \& et~al.}]{El-BadryQuataert2019}
{El-Badry}, K., {Quataert}, E., {Weisz}, D.~R., {Choksi}, N., \& et~al. 2019,
  \mnras, 482, 4528, \dodoi{10.1093/mnras/sty3007}

\bibitem[{{Fragione} {et~al.}(2018{\natexlab{a}}){Fragione}, {Ginsburg}, \&
  {Kocsis}}]{fragk18}
{Fragione}, G., {Ginsburg}, I., \& {Kocsis}, B. 2018{\natexlab{a}}, \apj, 856,
  92, \dodoi{10.3847/1538-4357/aab368}

\bibitem[{{Fragione} {et~al.}(2022){Fragione}, {Kocsis}, {Rasio}, \&
  {Silk}}]{FragioneKocsis2022}
{Fragione}, G., {Kocsis}, B., {Rasio}, F.~A., \& {Silk}, J. 2022, \apj, 927,
  231, \dodoi{10.3847/1538-4357/ac5026}

\bibitem[{{Fragione} \& {Leigh}(2018{\natexlab{a}})}]{fragle2018}
{Fragione}, G., \& {Leigh}, N. 2018{\natexlab{a}}, \mnras, 479, 3181,
  \dodoi{10.1093/mnras/sty1600}

\bibitem[{{Fragione} \& {Leigh}(2018{\natexlab{b}})}]{fragl2018b}
---. 2018{\natexlab{b}}, \mnras, 480, 5160, \dodoi{10.1093/mnras/sty2233}

\bibitem[{{Fragione} {et~al.}(2018{\natexlab{b}}){Fragione}, {Leigh},
  {Ginsburg}, \& {Kocsis}}]{fragleiginkoc18}
{Fragione}, G., {Leigh}, N. W.~C., {Ginsburg}, I., \& {Kocsis}, B.
  2018{\natexlab{b}}, \apj, 867, 119, \dodoi{10.3847/1538-4357/aae486}

\bibitem[{{Fragione} \& {Silk}(2020)}]{FragioneSilk2020}
{Fragione}, G., \& {Silk}, J. 2020, \mnras, 498, 4591,
  \dodoi{10.1093/mnras/staa2629}

\bibitem[{{Freitag} {et~al.}(2006){Freitag}, {G{\"u}rkan}, \&
  {Rasio}}]{frei2006}
{Freitag}, M., {G{\"u}rkan}, M.~A., \& {Rasio}, F.~A. 2006, \mnras, 368, 141,
  \dodoi{10.1111/j.1365-2966.2006.10096.x}

\bibitem[{{Fryer} {et~al.}(2001){Fryer}, {Woosley}, \& {Heger}}]{fryer2001}
{Fryer}, C.~L., {Woosley}, S.~E., \& {Heger}, A. 2001, \apj, 550, 372,
  \dodoi{10.1086/319719}

\bibitem[{{Furlong} {et~al.}(2015){Furlong}, {Bower}, {Theuns}, {Schaye}, \&
  et~al.}]{FurlongBower2015}
{Furlong}, M., {Bower}, R.~G., {Theuns}, T., {Schaye}, J., \& et~al. 2015,
  \mnras, 450, 4486, \dodoi{10.1093/mnras/stv852}

\bibitem[{{Gair} {et~al.}(2011){Gair}, {Mandel}, {Miller}, \&
  {Volonteri}}]{gair2011}
{Gair}, J.~R., {Mandel}, I., {Miller}, M.~C., \& {Volonteri}, M. 2011, General
  Relativity and Gravitation, 43, 485, \dodoi{10.1007/s10714-010-1104-3}

\bibitem[{{Georgiev} \& {B{\"o}ker}(2014)}]{GeorgievBoker2014}
{Georgiev}, I.~Y., \& {B{\"o}ker}, T. 2014, \mnras, 441, 3570,
  \dodoi{10.1093/mnras/stu797}

\bibitem[{{Georgiev} {et~al.}(2016){Georgiev}, {B{\"o}ker}, {Leigh},
  {L{\"u}tzgendorf}, \& {Neumayer}}]{georg2016}
{Georgiev}, I.~Y., {B{\"o}ker}, T., {Leigh}, N., {L{\"u}tzgendorf}, N., \&
  {Neumayer}, N. 2016, \mnras, 457, 2122, \dodoi{10.1093/mnras/stw093}

\bibitem[{{Giersz} {et~al.}(2015){Giersz}, {Leigh}, {Hypki}, {L\"{u}tzgendorf},
  \& {Askar}}]{gie15}
{Giersz}, M., {Leigh}, N.~W., {Hypki}, A., {L\"{u}tzgendorf}, N., \& {Askar},
  A. 2015, \mnras, 454, 3150, \dodoi{10.1093/mnras/stv2162}

\bibitem[{{Gonz{\'a}lez} {et~al.}(2021){Gonz{\'a}lez}, {Kremer}, {Chatterjee},
  {Fragione}, \& et~al.}]{GonzalezKremer2021}
{Gonz{\'a}lez}, E., {Kremer}, K., {Chatterjee}, S., {Fragione}, G., \& et~al.
  2021, \apjl, 908, L29, \dodoi{10.3847/2041-8213/abdf5b}

\bibitem[{{Gratton} {et~al.}(2003){Gratton}, {Bragaglia}, {Carretta},
  {Clementini}, \& et~al.}]{GrattonBragaglia2003}
{Gratton}, R.~G., {Bragaglia}, A., {Carretta}, E., {Clementini}, G., \& et~al.
  2003, \aap, 408, 529, \dodoi{10.1051/0004-6361:20031003}

\bibitem[{{Gratton} {et~al.}(1997){Gratton}, {Fusi Pecci}, {Carretta},
  {Clementini}, \& et~al.}]{GrattonFusiPecci1997}
{Gratton}, R.~G., {Fusi Pecci}, F., {Carretta}, E., {Clementini}, G., \& et~al.
  1997, \apj, 491, 749, \dodoi{10.1086/304987}

\bibitem[{{Greene} {et~al.}(2020){Greene}, {Strader}, \&
  {Ho}}]{GreeneStrader2020}
{Greene}, J.~E., {Strader}, J., \& {Ho}, L.~C. 2020, \araa, 58, 257,
  \dodoi{10.1146/annurev-astro-032620-021835}

\bibitem[{{G{\"u}rkan} {et~al.}(2004){G{\"u}rkan}, {Freitag}, \&
  {Rasio}}]{gurk2004}
{G{\"u}rkan}, M.~A., {Freitag}, M., \& {Rasio}, F.~A. 2004, \apj, 604, 632,
  \dodoi{10.1086/381968}

\bibitem[{{Hild} {et~al.}(2011){Hild}, {Abernathy}, {Acernese}, {Amaro-Seoane},
  \& et~al.}]{HildAbernathy2011}
{Hild}, S., {Abernathy}, M., {Acernese}, F., {Amaro-Seoane}, P., \& et~al.
  2011, Classical and Quantum Gravity, 28, 094013,
  \dodoi{10.1088/0264-9381/28/9/094013}

\bibitem[{{Hofmann} {et~al.}(2016){Hofmann}, {Barausse}, \&
  {Rezzolla}}]{HofmannBarausse2016}
{Hofmann}, F., {Barausse}, E., \& {Rezzolla}, L. 2016, \apjl, 825, L19,
  \dodoi{10.3847/2041-8205/825/2/L19}

\bibitem[{{Hopman} \& {Alexander}(2006)}]{hopale2006}
{Hopman}, C., \& {Alexander}, T. 2006, \apjl, 645, L133, \dodoi{10.1086/506273}

\bibitem[{{Hurley} {et~al.}(2000){Hurley}, {Pols}, \& {Tout}}]{HurleyPols2000}
{Hurley}, J.~R., {Pols}, O.~R., \& {Tout}, C.~A. 2000, \mnras, 315, 543,
  \dodoi{10.1046/j.1365-8711.2000.03426.x}

\bibitem[{{Jani} {et~al.}(2020){Jani}, {Shoemaker}, \&
  {Cutler}}]{JaniShoemaker2020}
{Jani}, K., {Shoemaker}, D., \& {Cutler}, C. 2020, Nature Astronomy, 4, 260,
  \dodoi{10.1038/s41550-019-0932-7}

\bibitem[{{Jim{\'e}nez-Forteza} {et~al.}(2017){Jim{\'e}nez-Forteza}, {Keitel},
  {Husa}, {Hannam}, \& et~al.}]{Jimenez-FortezaKeitel2017}
{Jim{\'e}nez-Forteza}, X., {Keitel}, D., {Husa}, S., {Hannam}, M., \& et~al.
  2017, \prd, 95, 064024, \dodoi{10.1103/PhysRevD.95.064024}

\bibitem[{{Kaaret} {et~al.}(2017){Kaaret}, {Feng}, \&
  {Roberts}}]{kaaret2017ARA&A..55..303K}
{Kaaret}, P., {Feng}, H., \& {Roberts}, T.~P. 2017, \araa, 55, 303,
  \dodoi{10.1146/annurev-astro-091916-055259}

\bibitem[{{Kocsis} {et~al.}(2011){Kocsis}, {Yunes}, \& {Loeb}}]{koc11}
{Kocsis}, B., {Yunes}, N., \& {Loeb}, A. 2011, \prd, 84, 024032,
  \dodoi{10.1103/PhysRevD.84.024032}

\bibitem[{{Kroupa}(2001)}]{kro01}
{Kroupa}, P. 2001, \mnras, 322, 231, \dodoi{10.1046/j.1365-8711.2001.04022.x}

\bibitem[{{Lin} {et~al.}(2018){Lin}, {Strader}, {Carrasco}, {Page},
  {Romanowsky}, {Homan}, {Irwin}, {Remillard}, {Godet}, {Webb}, {Baumgardt},
  {Wijnands}, {Barret}, {Duc}, {Brodie}, \& {Gwyn}}]{lin2018}
{Lin}, D., {Strader}, J., {Carrasco}, E.~R., {et~al.} 2018, Nature Astronomy,
  2, 656, \dodoi{10.1038/s41550-018-0493-1}

\bibitem[{{Loeb} \& {Rasio}(1994)}]{loeb1994}
{Loeb}, A., \& {Rasio}, F.~A. 1994, \apj, 432, 52, \dodoi{10.1086/174548}

\bibitem[{{Lousto} {et~al.}(2010){Lousto}, {Campanelli}, {Zlochower}, \&
  {Nakano}}]{lou10}
{Lousto}, C.~O., {Campanelli}, M., {Zlochower}, Y., \& {Nakano}, H. 2010,
  Classical and Quantum Gravity, 27, 114006,
  \dodoi{10.1088/0264-9381/27/11/114006}

\bibitem[{{Lousto} \& {Zlochower}(2008)}]{lou08}
{Lousto}, C.~O., \& {Zlochower}, Y. 2008, \prd, 77, 044028,
  \dodoi{10.1103/PhysRevD.77.044028}

\bibitem[{{Lousto} {et~al.}(2012){Lousto}, {Zlochower}, {Dotti}, \&
  {Volonteri}}]{lou12}
{Lousto}, C.~O., {Zlochower}, Y., {Dotti}, M., \& {Volonteri}, M. 2012, \prd,
  85, 084015, \dodoi{10.1103/PhysRevD.85.084015}

\bibitem[{{Luo} {et~al.}(2016){Luo}, {Chen}, {Duan}, {Gong}, \&
  et~al.}]{LuoChen2016}
{Luo}, J., {Chen}, L.-S., {Duan}, H.-Z., {Gong}, Y.-G., \& et~al. 2016,
  Classical and Quantum Gravity, 33, 035010,
  \dodoi{10.1088/0264-9381/33/3/035010}

\bibitem[{{MacLeod} {et~al.}(2016){MacLeod}, {Guillochon}, {Ramirez-Ruiz},
  {Kasen}, \& et~al.}]{MacLeodGuillochon2016}
{MacLeod}, M., {Guillochon}, J., {Ramirez-Ruiz}, E., {Kasen}, D., \& et~al.
  2016, \apj, 819, 3, \dodoi{10.3847/0004-637X/819/1/3}

\bibitem[{{Madau} \& {Fragos}(2017)}]{MadauFragos2017}
{Madau}, P., \& {Fragos}, T. 2017, \apj, 840, 39,
  \dodoi{10.3847/1538-4357/aa6af9}

\bibitem[{{Madau} \& {Rees}(2001)}]{madau2001}
{Madau}, P., \& {Rees}, M.~J. 2001, \apjl, 551, L27, \dodoi{10.1086/319848}

\bibitem[{{Mandel} {et~al.}(2008){Mandel}, {Brown}, {Gair}, \&
  {Miller}}]{MandelBrown2008}
{Mandel}, I., {Brown}, D.~A., {Gair}, J.~R., \& {Miller}, M.~C. 2008, \apj,
  681, 1431, \dodoi{10.1086/588246}

\bibitem[{{Mapelli} {et~al.}(2021){Mapelli}, {Dall'Amico}, {Bouffanais},
  {Giacobbo}, \& et~al.}]{MapelliDall'Amico2021}
{Mapelli}, M., {Dall'Amico}, M., {Bouffanais}, Y., {Giacobbo}, N., \& et~al.
  2021, \mnras, 505, 339, \dodoi{10.1093/mnras/stab1334}

\bibitem[{{McKernan} {et~al.}(2014){McKernan}, {Ford}, {Kocsis}, {Lyra}, \&
  {Winter}}]{McKernan+2014}
{McKernan}, B., {Ford}, K.~E.~S., {Kocsis}, B., {Lyra}, W., \& {Winter}, L.~M.
  2014, \mnras, 441, 900, \dodoi{10.1093/mnras/stu553}

\bibitem[{{McKernan} {et~al.}(2012){McKernan}, {Ford}, {Lyra}, \&
  {Perets}}]{McKernan+2012}
{McKernan}, B., {Ford}, K.~E.~S., {Lyra}, W., \& {Perets}, H.~B. 2012, \mnras,
  425, 460, \dodoi{10.1111/j.1365-2966.2012.21486.x}

\bibitem[{{Mi{\'c}i{\'c}} {et~al.}(2022){Mi{\'c}i{\'c}}, {Irwin}, \&
  {Lin}}]{MicicIrwin2022}
{Mi{\'c}i{\'c}}, M., {Irwin}, J.~A., \& {Lin}, D. 2022, arXiv e-prints,
  arXiv:2203.10136.
\newblock \doarXiv{2203.10136}

\bibitem[{{Miller}(2002)}]{Miller2002}
{Miller}, M.~C. 2002, \apj, 581, 438, \dodoi{10.1086/344156}

\bibitem[{{Miller} \& {Hamilton}(2002)}]{mil02b}
{Miller}, M.~C., \& {Hamilton}, D.~P. 2002, \mnras, 330, 232,
  \dodoi{10.1046/j.1365-8711.2002.05112.x}

\bibitem[{{Natarajan}(2021)}]{Natarajan2021}
{Natarajan}, P. 2021, \mnras, 501, 1413, \dodoi{10.1093/mnras/staa3724}

\bibitem[{{Neumayer} {et~al.}(2020){Neumayer}, {Seth}, \&
  {B{\"o}ker}}]{NeumayerSeth2020}
{Neumayer}, N., {Seth}, A., \& {B{\"o}ker}, T. 2020, \aapr, 28, 4,
  \dodoi{10.1007/s00159-020-00125-0}

\bibitem[{{O'Leary} {et~al.}(2016){O'Leary}, {Meiron}, \& {Kocsis}}]{omk16}
{O'Leary}, R.~M., {Meiron}, Y., \& {Kocsis}, B. 2016, \apjl, 824, L12,
  \dodoi{10.3847/2041-8205/824/1/L12}

\bibitem[{{O'Leary} {et~al.}(2006){O'Leary}, {Rasio}, {Fregeau}, {Ivanova}, \&
  {O'Shaughnessy}}]{OLeary+2006}
{O'Leary}, R.~M., {Rasio}, F.~A., {Fregeau}, J.~M., {Ivanova}, N., \&
  {O'Shaughnessy}, R. 2006, \apj, 637, 937, \dodoi{10.1086/498446}

\bibitem[{{Pan} {et~al.}(2012){Pan}, {Loeb}, \& {Kasen}}]{panloeb2012}
{Pan}, T., {Loeb}, A., \& {Kasen}, D. 2012, \mnras, 423, 2203,
  \dodoi{10.1111/j.1365-2966.2012.21030.x}

\bibitem[{{Pechetti} {et~al.}(2022){Pechetti}, {Seth}, {Kamann}, {Caldwell}, \&
  et~al.}]{PechettiSeth2022}
{Pechetti}, R., {Seth}, A., {Kamann}, S., {Caldwell}, N., \& et~al. 2022, \apj,
  924, 48, \dodoi{10.3847/1538-4357/ac339f}

\bibitem[{{Peng} {et~al.}(2019){Peng}, {Yang}, {Shen}, {Wang}, {Zou}, \&
  {Zhang}}]{peng2019}
{Peng}, Z.-K., {Yang}, Y.-S., {Shen}, R.-F., {et~al.} 2019, \apjl, 884, L34,
  \dodoi{10.3847/2041-8213/ab481b}

\bibitem[{{Planck Collaboration} {et~al.}(2016){Planck Collaboration}, {Ade},
  {Aghanim}, {Arnaud}, \& et~al.}]{PlanckCollaborationAde2016}
{Planck Collaboration}, {Ade}, P.~A.~R., {Aghanim}, N., {Arnaud}, M., \& et~al.
  2016, \aap, 594, A13, \dodoi{10.1051/0004-6361/201525830}

\bibitem[{{Portegies Zwart} \& {McMillan}(2002)}]{por02}
{Portegies Zwart}, S.~F., \& {McMillan}, S.~L.~W. 2002, \apj, 576, 899,
  \dodoi{10.1086/341798}

\bibitem[{{Rasskazov} {et~al.}(2020){Rasskazov}, {Fragione}, \&
  {Kocsis}}]{RasskazovFragione2020}
{Rasskazov}, A., {Fragione}, G., \& {Kocsis}, B. 2020, \apj, 899, 149,
  \dodoi{10.3847/1538-4357/aba2f4}

\bibitem[{Robson {et~al.}(2019)Robson, Cornish, \& Liu}]{robson2019}
Robson, T., Cornish, N.~J., \& Liu, C. 2019, Classical and Quantum Gravity, 36,
  105011, \dodoi{10.1088/1361-6382/ab1101}

\bibitem[{{Rose} {et~al.}(2021){Rose}, {Naoz}, {Sari}, \&
  {Linial}}]{RoseNaoz2021}
{Rose}, S.~C., {Naoz}, S., {Sari}, R., \& {Linial}, I. 2021, arXiv e-prints,
  arXiv:2201.00022.
\newblock \doarXiv{2201.00022}

\bibitem[{{Rosswog} {et~al.}(2008){Rosswog}, {Ramirez-Ruiz}, \&
  {Hix}}]{rossw2008}
{Rosswog}, S., {Ramirez-Ruiz}, E., \& {Hix}, W.~R. 2008, \apj, 679, 1385,
  \dodoi{10.1086/528738}

\bibitem[{{Rosswog} {et~al.}(2009){Rosswog}, {Ramirez-Ruiz}, \&
  {Hix}}]{RosswogRamirez-Ruiz2009}
---. 2009, \apj, 695, 404, \dodoi{10.1088/0004-637X/695/1/404}

\bibitem[{{Shen}(2019)}]{shen2019}
{Shen}, R.-F. 2019, \apjl, 871, L17, \dodoi{10.3847/2041-8213/aafc64}

\bibitem[{{Silk}(2017)}]{Silk2017}
{Silk}, J. 2017, \apjl, 839, L13, \dodoi{10.3847/2041-8213/aa67da}

\bibitem[{{Stone} {et~al.}(2017){Stone}, {K{\"u}pper}, \&
  {Ostriker}}]{StoneKupper2017}
{Stone}, N.~C., {K{\"u}pper}, A. H.~W., \& {Ostriker}, J.~P. 2017, \mnras, 467,
  4180, \dodoi{10.1093/mnras/stx097}

\bibitem[{{Stone} \& {Ostriker}(2015)}]{StoneOstriker2015}
{Stone}, N.~C., \& {Ostriker}, J.~P. 2015, \apjl, 806, L28,
  \dodoi{10.1088/2041-8205/806/2/L28}

\bibitem[{{Tagawa} {et~al.}(2019){Tagawa}, {Haiman}, \& {Kocsis}}]{tagawa2019}
{Tagawa}, H., {Haiman}, Z., \& {Kocsis}, B. 2019, arXiv e-prints,
  arXiv:1909.10517.
\newblock \doarXiv{1909.10517}

\bibitem[{{Tagawa} {et~al.}(2020){Tagawa}, {Haiman}, \&
  {Kocsis}}]{TagawaHaiman2020}
---. 2020, \apj, 892, 36, \dodoi{10.3847/1538-4357/ab7922}

\bibitem[{{The LIGO Scientific Collaboration} \& {the Virgo
  Collaboration}(2020)}]{ligo2020new2}
{The LIGO Scientific Collaboration}, \& {the Virgo Collaboration}. 2020, \apjl,
  900, L13, \dodoi{10.3847/2041-8213/aba493}

\bibitem[{{VandenBerg} {et~al.}(2013){VandenBerg}, {Brogaard}, {Leaman}, \&
  {Casagrande}}]{VandenBergBrogaard2013}
{VandenBerg}, D.~A., {Brogaard}, K., {Leaman}, R., \& {Casagrande}, L. 2013,
  \apj, 775, 134, \dodoi{10.1088/0004-637X/775/2/134}

\bibitem[{{Vogelsberger} {et~al.}(2014){Vogelsberger}, {Genel}, {Springel},
  {Torrey}, \& et~al.}]{VogelsbergerGenel2014}
{Vogelsberger}, M., {Genel}, S., {Springel}, V., {Torrey}, P., \& et~al. 2014,
  \mnras, 444, 1518, \dodoi{10.1093/mnras/stu1536}

\bibitem[{{Weatherford} {et~al.}(2021){Weatherford}, {Fragione}, {Kremer},
  {Chatterjee}, \& et~al.}]{WeatherfordFragione2021}
{Weatherford}, N.~C., {Fragione}, G., {Kremer}, K., {Chatterjee}, S., \& et~al.
  2021, \apjl, 907, L25, \dodoi{10.3847/2041-8213/abd79c}

\bibitem[{{Yagi} \& {Seto}(2011)}]{YagiSeto2011}
{Yagi}, K., \& {Seto}, N. 2011, \prd, 83, 044011,
  \dodoi{10.1103/PhysRevD.83.044011}

\end{thebibliography}

\end{document}